\documentclass[twocolumn,reprint]{revtex4}
\usepackage{graphicx}
\usepackage{amsmath}
\usepackage{wasysym}
\usepackage{setspace}

\begin{document}


\title{Nanophotonic Luminescent Solar Concentrators}

\author{I. Rousseau and V. Wood}

\address{Laboratory for Nanoelectronics\\Department of Information Technology and Electrical Engineering\\ETH Zurich\\Gloriastrasse 35\\CH-8092 Zurich, Switzerland}

\email{vwood@ethz.ch}

\begin{abstract}

We investigate the connection between photonic local density of states and luminescent solar concentrator (LSC) performance in two manufacturable nanocavity LSC structures, a bilayer slab and a slab photonic crystal. Finite-difference time-domain electromagnetic simulations show that the waveguided luminescence photon flux can be enhanced up to 30\% for the photonic crystal design over a conventional LSC operating in the ray optic limit assuming the same number of excited lumophores.  Further photonic engineering could realize an increase of up to one order of magnitude in the flux of waveguided luminescence.

\end{abstract}

\maketitle

The luminescent solar concentrator (LSC) could decrease the installed cost of solar energy through building integration \cite{Debije2011}.  The LSC, a semi-transparent waveguide with embedded lumophores, concentrates sunlight by frequency downconversion; the lumophores absorb diffuse incident sunlight and luminesce at a redder, Stokes-shifted wavelength.  The majority of the luminescence is emitted into modes that can be guided by total internal reflection (TIR) to the waveguide edges, upon which small-area, high efficiency solar cells are normally fastened.  

Despite the LSC's simplicity, the concept has not been commercialized due to low performance.  Experimental realizations have demonstrated a twelve-fold concentration of solar flux \cite{Currie2008} and power conversion efficiency of 7.2\%, well below the theoretical predictions of a flux concentration in excess of 100 \cite{Smestad1990} and power conversion efficiency of 26.8\% \cite{Markvart2006}.  Reabsorption of luminescence and subsequent re-emission into non-waveguided modes has been identified as the primary performance bottleneck  \cite{Batchelder1979,Debije2008,Olson1981,Weber1976,Wilson2010}.  

\begin{figure}[h!tb]
	\centering
		\includegraphics[width=0.425\textwidth]{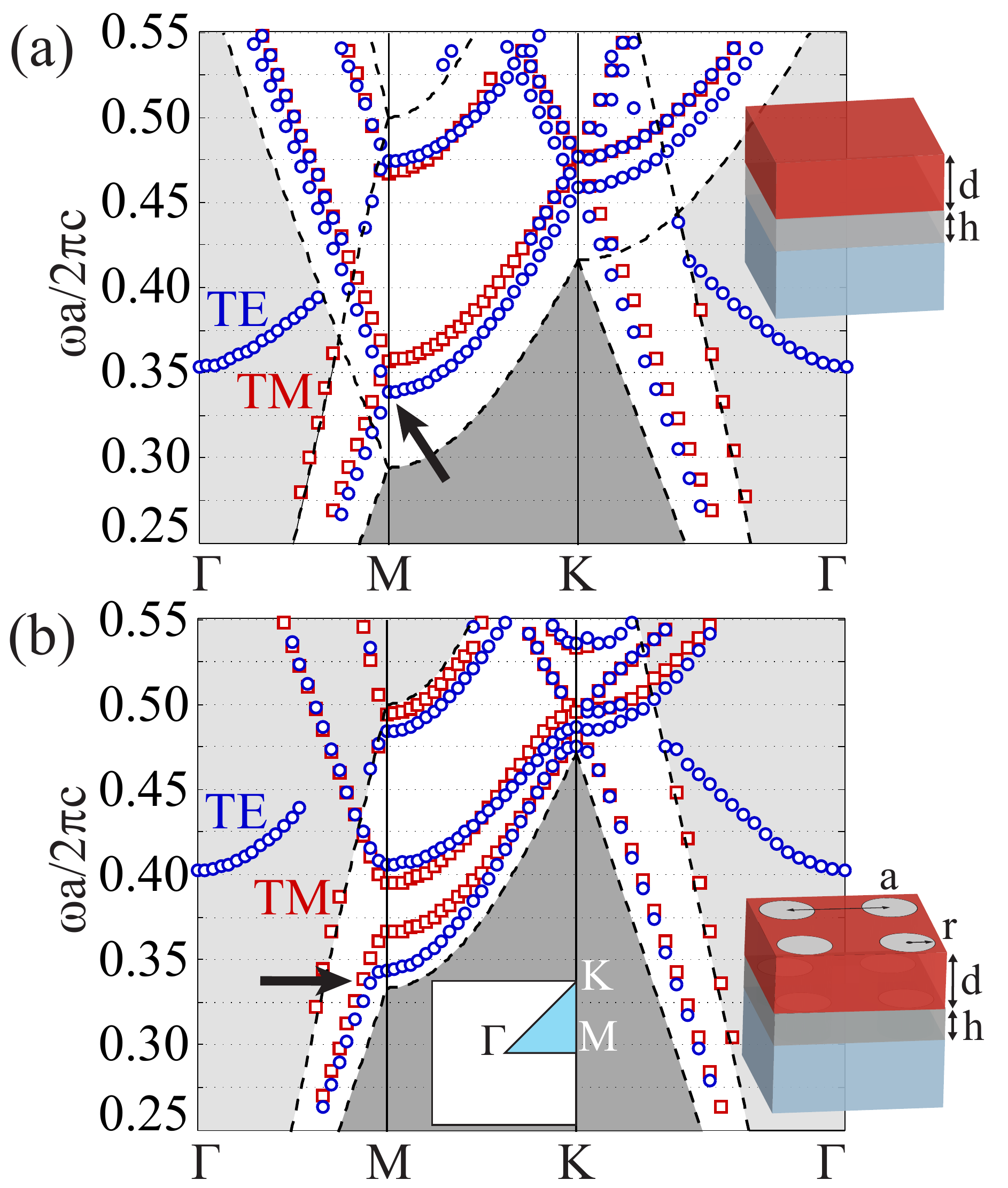}
	\caption{The band structures for the (a) bilayer slab and (b) slab photonic crystal LSCs.  Blue circles represent TE-like modes and red squares, TM-like modes.  The white region corresponds to generalized TIR modes, and the light gray represents modes that can leak into air.  The discrete periodicity of the PC-LSC lifts the degeneracies at the $M$-point, as indicated by the black arrow.  The insets show the nanocavity LSC design, a waveguide core (red, $n=1.7$) coupled through a cladding (gray, $n=1.3$) to the underlying substrate (blue).}
	\label{fig:one}
\end{figure}

While prior work has demonstrated that LSCs consisting of lumophores embedded in optical nanocavities exhibit enhanced waveguiding \cite{Gutmann2012} and reduced reabsorption \cite{Giebink2011}, the nanocavity modifies the photonic local density of states (LDOS) and therefore the spatial and temporal luminescence distributations \cite{John1987,Yablonovitch1987}.  Here, we investigate the effect of the modified LDOS on LSC performance using first-principles simulations of Maxwell's equations in two realistic nanocavity LSC designs (Fig. \ref{fig:one}, insets).  After establishing a link between photonic LDOS and LSC performance, we use finite difference time domain (FDTD) simulations to show that a nanocavity LSC can increase the flux of waveguided luminescence photons by up to 30\% over a conventional LSC.  Finally, we assess the maximum theoretical performance gains from LDOS engineering in the LSC.

First, we define a  metric of LSC performance, connect the performance metric to the photonic LDOS, and determine the conditions under which LSC performance comparisons can be made on the basis of the photonic LDOS.  Luminescence photons are spontaneously emitted into one of many photonic modes of the LSC.  These can be divided into two groups based on the wavevector in the LSC plane ($\textbf{k}_{\parallel}$): non-waveguided ($\omega \ge  c |\textbf{k}_{\parallel}|$) and total internally reflected (TIR, $\omega <  c |\textbf{k}_{\parallel}|$) modes.  In conventional LSCs, lumophore-filled waveguides of refractive index $n$ much thicker than the luminescence wavelength, the fraction $f_{tir}$
\begin{equation}
	f_{tir} = \sqrt{n^2 - 1} / n
	\label{eq:ftir}
\end{equation}
of photonic modes corresponds to TIR modes \cite{Debije2011}.  According to ray tracing simulations, photons emitted into non-waveguided modes are lost into air after multiple reflections \cite{Batchelder1979}.  Thus, our analysis assumes that only luminescence photons emitted into TIR modes can be collected by the solar cells attached to the LSC edges.

Assuming the lumophore has unity quantum yield and the LSC surfaces are smooth, a photon emitted into a TIR mode can either be re-absorbed by the lumophore or absorbed by the solar cells affixed to the LSC edges \cite{Yablonovitch1980,Markvart2006}. Starting from the quantum optical master equation for a single, lossy photonic mode weakly coupled to a lumophore, we derive a recursion relation for the occupation probability ($\rho_n$) of a $n$-photon Fock state \cite{supplemental}
\begin{eqnarray}
   \dot{\rho}_n &=& A_{\omega k} n \rho_{n-1} - A_{\omega k} (n + 1) \rho_{n} - B_{\omega k} n \rho_n  \nonumber \\
       &-& D_{\omega k} n \rho_n + (n+1) (B_{\omega k} + D_{\omega k}) \rho_{n+1}   \label{eq:recursion}
\end{eqnarray}
Here, $A_{\omega k}$ and $B_{\omega k}$ are the photon emission and absorption rates from Fermi's Golden Rule \cite{Abraham1977}, and $D_{\omega k}$ is the rate at which photons leak out of the photonic mode \cite{Breuer2007}, which, in the case of the LSC, we assume is the rate at which luminescence photons are collected by solar cells affixed to the LSC edges.

The solution to Eq. \ref{eq:recursion} is a geometric series \cite{supplemental}, and the steady-state rate at which photons are collected by the attached solar cells from a single TIR mode is
\begin{equation}
	D_{\omega k} \bar{n} = D_{\omega k} \sum_n n \bar{\rho}_n  = \frac{ A_{\omega k} D_{\omega k} }{B_{\omega k} - A_{\omega k} + D_{\omega k}}
	\label{eq:detailedBalanceSolution}
\end{equation}
Eq. \ref{eq:detailedBalanceSolution} is the emission rate divided by the sum of the reabsorption and collection rates minus the emission rate.  If photons are extracted by the solar cells much faster than they can be reabsorbed ($D_{\omega k} \gg B_{\omega k}$), then Eq. \ref{eq:detailedBalanceSolution} can be further simplified such that the photon collection rate for a single mode is then equal to the emission rate, $	D_{\omega k} \bar{n}_{\omega k} = A_{\omega k} $.

Experimentally, this regime corresponds to either small LSCs or LSCs containing lumophores with small reabsorption, which has been demonstrated by exploiting F\"{o}rster resonant energy transfer or intersystem crossing \cite{Currie2008}.  Eq. \ref{eq:detailedBalanceSolution} must be summed over all wavevectors corresponding to TIR modes ($|\textbf{k}_{\parallel}| > \omega / c$)  to find the total flux collected by attached solar cells at each frequency.  In this operation regime, LSCs with the same species and number of excited lumophores can be directly compared on the basis of TIR LDOS \cite{supplemental}.  Here, we compare TIR LDOS in different LSC designs: a ``conventional" LSC comprised of lumophores dispersed in a thick dielectric waveguide and two different nanophotonic LSCs with lumophores embedded in a nanocavity.

We calculate the photonic LDOS for two realistic nanocavity LSC designs, each with a waveguide core comprised of the organic lumophore DCM$_2$-doped Alq$_3$ ($n=1.7$) and a fluoropolymer ($n=1.3$) waveguide cladding.  The nanocavity formed by the air-core-cladding layers is coupled to the underlying waveguide-substrate for transport of luminescence to attached solar cells.  In the bilayer slab LSC (Fig. \ref{fig:one}a, inset), the waveguide is coupled through the cladding layer to the underlying flint glass ($n=1.7$) substrate-waveguide \cite{Giebink2011}.  The second design, a slab photonic crystal (PC) LSC (Fig. \ref{fig:one}b, inset) has the same dimensions as the bilayer slab but contains a square lattice of air holes extruded through the core and cladding layers.  Since the air holes decrease the core's average refractive index, the substrate-waveguide refractive index is reduced to that of crown glass ($n=1.5$), which would reduce material costs.

The photonic LDOS in three dimensions for each structure is computed using a freely available finite-difference time-domain (FDTD) software package \cite{Oskooi2010}.  The FDTD method is chosen because the LDOS can be calculated over a broad frequency range in a single simulation \cite{Koenderink2006}. To simulate an infinite periodic photonic crystal, Bloch periodic boundary conditions in the in-plane directions and absorbing boundary conditions in the vertical directions are selected.
In order to calculate the LDOS, the electric field transient is recorded at a point after excitation by a point broadband Gaussian current source ($\textbf{p}(\textbf{r},\omega)$) at that same location.  The Fourier transform of the electric field transient normalized by the excitation pulse spectrum yields the photonic local density of states for a given location and source orientation  \cite{Oskooi2013}
\begin{equation}
	B(\omega,\textbf{r},\textbf{d}) = - \frac{2}{\pi} n(\textbf{r})^2 \frac{\textrm{Re} \big{[} \textbf{E}(\textbf{r},\omega) \cdot \textbf{p}^{\ast}(\textbf{r},\omega) \big{]} }{|\textbf{p}(\textbf{r},\omega)|^2}
	\label{eq:fdtdldos}
\end{equation}
Since the LDOS depends on the dipole location and orientation \cite{Lee2000}, LDOS calculations are carried out for thirty randomly selected locations within the waveguide core volume, three Cartesian dipole orientations, and for 2401 wavevectors in a rectangular mesh spanning the irreducible Brillouin zone of the square lattice.  Further details explaining the FDTD simulations, $\textbf{k}$-space integration, and LDOS normalization are found in the supplementary material \cite{supplemental}.



Fig. \ref{fig:one} displays the simulated band structures, taken from local maxima in the LDOS, as a function of in-plane wavevector ($\textbf{k}_{\parallel}$).  The blue circles and red squares indicate TE-like ($\textbf{E} \parallel \textbf{k}_{\parallel}$) and TM-like ($\textbf{E} \perp \textbf{k}_{\parallel}$) modes, respectively.  The white region corresponds to generalized TIR modes in the substrate-waveguide \cite{Joannopoulos2008}.  Luminescence emitted into these modes can be transported within the lumophore-free substrate-waveguide to the attached solar cells.

The simulated dimensions, $h/a=0.9$, $d/a=0.5$, and $r/a=0.275$, while not optimized, are selected to satisfy several criteria.  Comparing the two bandstructures, we see that the PC-LSC's discrete translational symmetry splits the TE- and TM-like guided photonic bands at the $M$-point in the band structure (black arrow).  To enhance emission into TIR modes, the PC-LSC dimensions are chosen such that the lumophore photoluminescence spectrum overlaps with Van Hove singularities for the dielectric (valence) bands residing in the white generalized TIR region. Van Hove singularities occur at saddle points in the dispersion relation (here, the $M$-point) and result in peaks in the photonic DOS \cite{Busch1998,Joannopoulos2008}.  Finally, modes below the light line in the substrate-waveguide (dark gray) should be eliminated so that luminescence is not trapped in the nanocavity.
These design considerations are opposite to those previously proposed for enhancement of light extraction in light-emitting diodes \cite{Boroditsky1999,Fan1997}.

\begin{figure}
	\centering
		\includegraphics[width=0.425\textwidth]{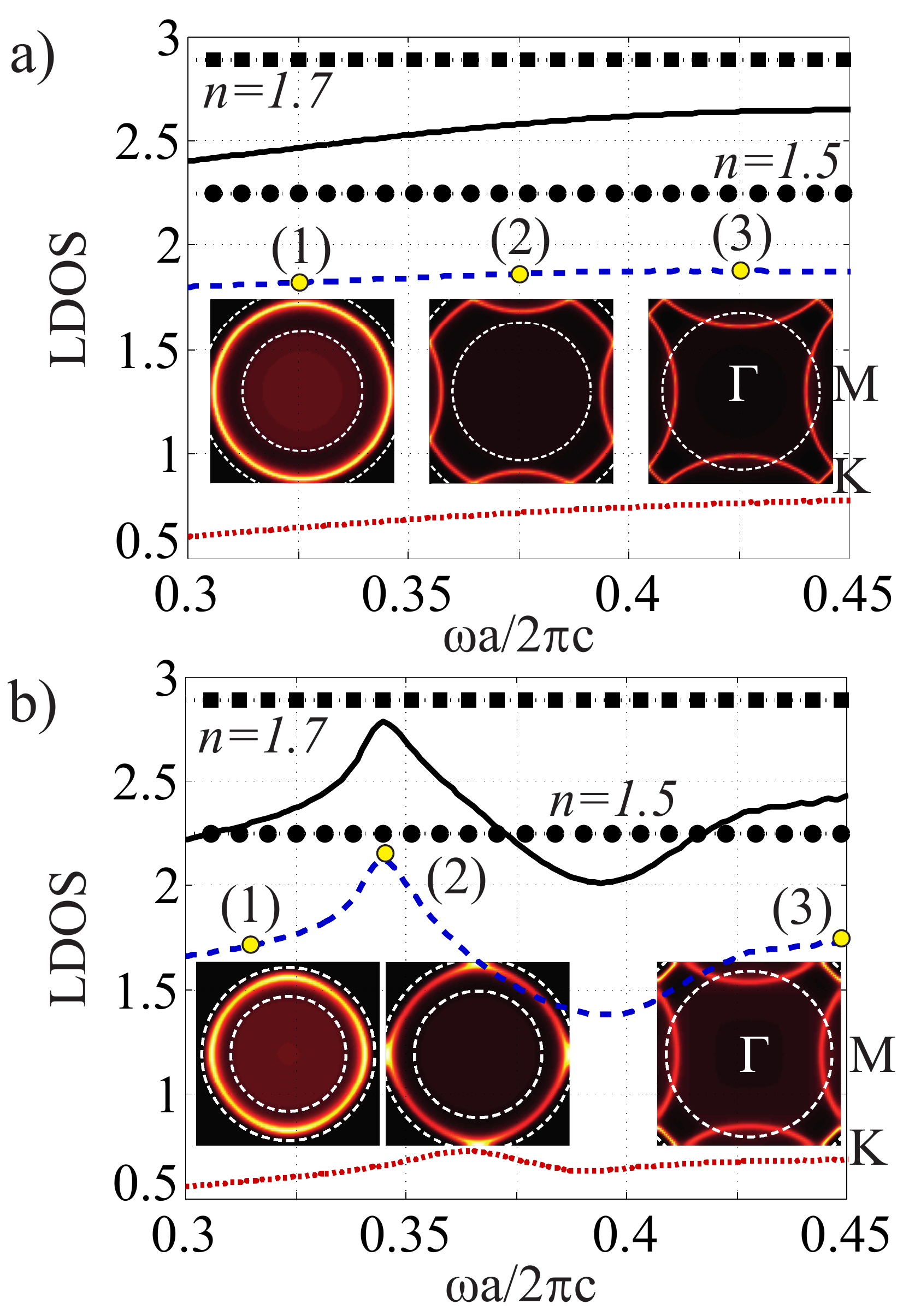}
	\caption{Total photonic LDOS (black solid lines) for the (a) bilayer LSC and (b) PC-LSC, normalized to the vacuum LDOS.  Blue dashed and red dotted lines indicate the contribution of TE-like and TM-like modes to the total LDOS, respectively.  Black squares and circles represent the LDOS for $n=1.7$ and $n=1.5$ conventional LSCs.  Insets show the distributions of spontaneous emission in $\textbf{k}_{\parallel}$-space at the designated frequencies, represented on a relative scale at each frequency.  The concentric white dashed circles represent the light lines in air and the underlying substrate, respectively.}
	\label{fig:two}
\end{figure}
Fig. \ref{fig:two} compares the total simulated LDOS normalized to the total vacuum LDOS.  The reference cases are conventional LSCs in which the lumophores are dispersed in a glass substrate that is much thicker than the wavelength of the emitted light.  The bilayer slab (Fig.\ref{fig:two}a) LDOS is approximately 80\% of the $n=1.7$ conventional LSC, consistent with previous experimental and theoretical work on dielectric slabs \cite{Urbach1998}.   
The PC-LSC LDOS (Fig.\ref{fig:two}b) falls between that of $n=1.5$ and $n=1.7$ conventional LSCs, but does not exceed the $n=1.7$ conventional LSC \cite{Boroditsky1999}.  The TE-like (blue) and TM-like (red) contributions to the total LDOS confirm that the LDOS enhancement lies at the Van Hove singularity corresponding to the $M$-point (Fig. \ref{fig:two}b).  Finally, in the PC-LSC design, spontaneous emission is inhibited between $\omega a /2 \pi c = 0.38$ and 0.41 in the PC-LSC; fewer photons are emitted in the region of strongest overlap between lumophore absorption and luminescence spectra.

The insets in Fig. \ref{fig:two} provide visual confirmation that the embedment of lumophores in nanocavities results in directional spontaneous emission distributions.  The concentric white dashed circles indicate the light lines in air and in the substrate.  Luminescence emitted between these two circles is waveguided by generalized TIR.  In both nanocavity LSC designs, a ring corresponding to a single guided TE mode increases in size with frequency and is eventually folded about the band edges due to the Bloch periodic boundary conditions.  Since the bilayer LSC has continuous translational symmetry, the folding has no bearing on LSC performance.  The folding in the PC-LSC, on the other hand, begins to decrease the TIR LDOS starting at $\omega a / 2\pi c = 0.42$, where the second band crosses the light line in air (Fig. \ref{fig:one}b). However, through the choice of a larger lattice constant, the strong dielectric band emission at $\omega a / 2 \pi c = 0.344$ at the $M$-point can be utilized to increase the TIR LDOS.  By integrating these partial LDOS distributions between the two light lines, we compute the TIR LDOS for each structure.

\begin{figure}[h!tb]
	\centering
		\includegraphics[width=0.45\textwidth]{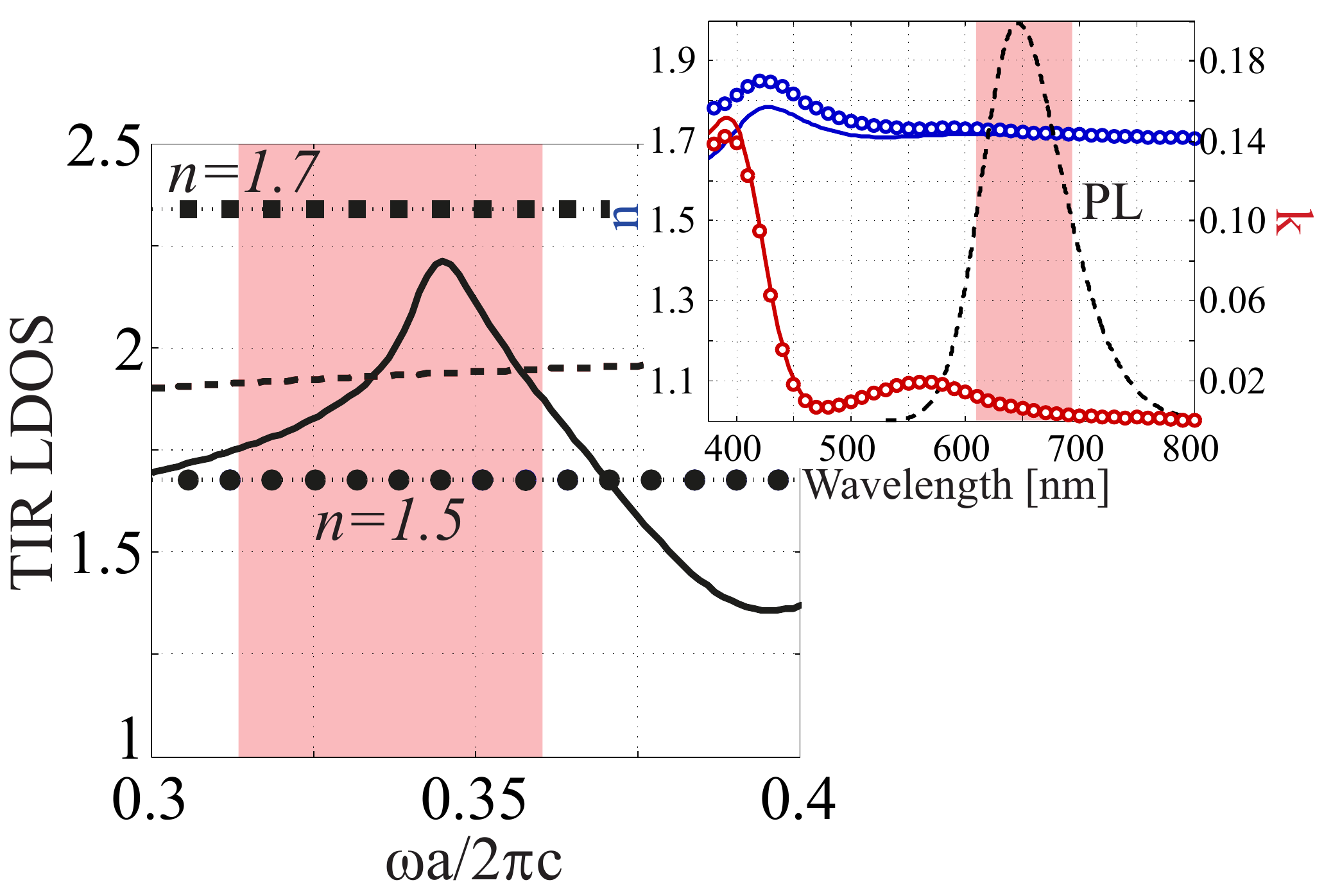}
	\caption{LDOS corresponding to TIR modes in the PC-LSC (solid line) and bilayer LSC (dashed) normalized to the total vacuum LDOS.  The black squares and circles indicate the TIR LDOS for $n=1.5$ and $n=1.7$ conventional LSCs.  The inset shows the measured real (blue) and imaginary (red) parts of the refractive index and photoluminescence spectrum (black dashed line) of a Alq$_3$:DCM$_2$ thin film \cite{supplemental}.  The pink shading delineates the FWHM of the photoluminescence spectrum for a specific lattice constant, $a=222$ nm.}
	\label{fig:three}
\end{figure}Fig. \ref{fig:three} compares the TIR LDOS for the nanocavity LSCs with those of conventional LSCs.  The nanocavity LSCs are compared to conventional LSCs with the same substrate refractive index.  The bilayer LSC's TIR LDOS is 15\% less than that of the $n=1.7$ reference.  The PC-LSC TIR LDOS exceeds that of a conventional LSC with $n=1.5$ by up to 31\% at the $M$-point.  We choose the lattice constant $a=222$ nm, which maximizes the product of the TIR LDOS with the Alq$_3$:DCM$_2$ photoluminescence spectrum (inset) when integrated over the luminescence photon frequency.  For this choice of $a$ (pink squares), the PC-LSC demonstrates a modest 10\% increase in the waveguided luminescent photon flux over the $n=1.5$ conventional LSC.

The maximum TIR LDOS enhancement of spontaneous emission is limited by lumophore photoluminescence spectrum linewidth ($Q_m=\omega_0/\Delta \omega$) to $Q_m/4 f_{tir}$ \cite{Boroditsky1999}.  LSC lumophore $Q_m$-factors range from three to seven for organics (Fig. \ref{fig:three}, inset) to thirty for state-of-the-art colloidal semicondutor quantum dots \cite{Bawendi2013}.  Thus, the maximum TIR LDOS enhancment is approximately ten.

In conclusion, we calculated the detailed spatial and spectral luminescence distributions for two nanophotonic LSC designs, which are chosen based on past work as well as manfuacturability considerations \cite{Yokoo2003}.  Although the directional luminescence distributions are desirable for minimizing reabsorption losses \cite{Giebink2011}, both the total and waveguided spontaneous emission rates are less than the bulk rate for the waveguide core material.  By patterning the waveguide core into a photonic crystal, we find that the spontaneous emission rate into guided modes of the bulk material can be recovered by utilizing dielectric band edge emission in the photonic crystal.  The inhibition of the total spontaneous emission rate in a photonic crystal could be further exploited to reduce reabsorption losses in the large LSC limit.



\section{Acknowledgements}
This research was supported by the Swiss National Science Foundation through the National Centre of Competence in Research Quantum Science and Technology and by ETH Research Grant ETH-42 12-2.  The authors would like to thank Atac Imamo\u{g}lu, Christian Hafner, the Computational Optics group at ETH, and Deniz Bozyigit for helpful discussions.

\end{document}